# Potential energy landscape description with Gamma distribution for supercooled liquids and glasses

Hongqin Liu

Integrated High Performance Computing, Shared Services Canada, Montreal, Canada.

The potential energy landscape (PEL) theory stands as one of the most successful frameworks for understanding supercooled liquids and glassy systems. A central element of this theory is the configurational entropy, $S_c$, which is traditionally represented by a symmetric Gaussian distribution. However, the asymmetric nature of the potential energy of inherent structures, $E_{IS}$, poses a challenge to such a representation across wide regions of configurational space. In addition, the Gaussian distribution fails to represent fragile to strong transition (FST) observed in various fluids. In this work, we demonstrate that an asymmetric distribution, specifically the Gamma distribution, provides effective description of both $S_c$ and $E_{IS}$ over broad ranges of density and temperature ($T$). The FST is interpreted through shifts of the $E_{IS}$ distribution and the curvature change of the $E_{IS} \sim 1/T$ relation. In terms of energy changes, the FST is comparable to a liquid-liquid phase transition. Moreover, the revised PEL framework yields an equation of state that incorporates a singular term diverging at a glassy or jammed state—an important feature for accurately describing the pressure behavior of these systems.

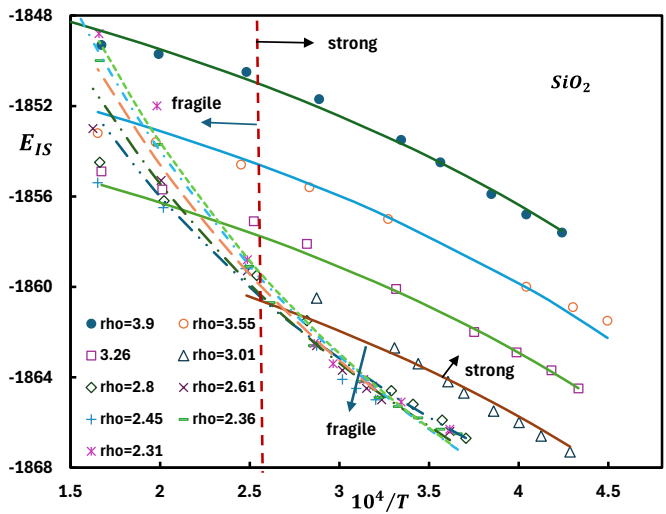

Emails: hongqin.liu@ssc-spc.gc.ca; hqliu2000@gmail.com.

# Introduction

Supercooled liquids and glassy materials exhibit complex dynamics that challenge our understanding of their thermodynamic and kinetic behaviours [1,2,3]. The potential energy landscape (PEL) framework has emerged as a powerful theoretical tool for describing the relationship between a system's microscopic configurations and its macroscopic properties [3,4,5,6,7]. By mapping the system's structure onto a multidimensional energy surface, the PEL theory provides insight into phenomena such as glass transitions, dynamical heterogeneity, and relaxation processes. Despite its successes, vast majority of existing PEL models rely on simplifying assumptions—most notably, the use of a Gaussian distribution to characterize the density of states associated with the local potential energy of the inherent structures known as basins [3,4,5,6,7]. While this assumption is computationally convenient, it imposes a symmetric potential energy distribution, which becomes increasingly inadequate as the configurational space broadens and the system's complexity grows, macroscopically caused by temperature or density changes.

An important observation from computer simulations is the fragile-to-strong transition (FST), also referred to as the fragility crossover. This phenomenon has been observed in typical network-structured fluids, such as $SiO_2$ [8,9,10]; hydrogen-bonded $H_2O$ [11,12]; as well as in "regular" fluids governed by the van der Waals forces, such as the Lennard-Jones fluids [6,13,14]. For strong liquids, the potential energy landscape is often considered relatively "simple," characterized by a narrow distribution of energy minima and relatively uniform energy barriers between them. Such a simplicity can be approximated by a Gaussian-like distribution for the potential energy of the inherent structures, as a Gaussian distribution suggests a relatively homogeneous material with minimal variations in the local structural environments.

Conversely, fragile liquids exhibit a much "rougher" and more complex PEL, with a broader, more heterogeneous distribution of energy minima and a wider variety of barrier heights. This underlying complexity gives rise to greater sensitivity in their dynamics and thermodynamics as the environment variables (temperature, volume) change. Consequently, it becomes evident that the Gaussian distribution is incapable to describe the FST effectively.

To overcome the limitations of current models, we propose a revised PEL theory that incorporates an asymmetric Gamma distribution for describing the energy state distribution of the systems interested in this work. This enhanced model offers a more flexible and accurate representation of configurational entropy, effectively capturing key behaviours across broad ranges of temperatures and densities.

We apply the revised PEL framework to a binary mixture of Lennard-Jones fluid (BMLJ), the $SiO_2$ system, and the q-TIP4P/F water system, demonstrating its capability to accurately predict the temperature dependence of the potential energy of the inherent structures and the configurational entropy. Notably, the model successfully captures the FST, indicated by the shifts of the potential energy distribution (from Gaussian to non-Gaussian) and curvature change of the potential energy - temperature relation.

Finally, the new PEL theory provides a new type of equation of state (EoS) for supercooled fluids and glasses. This EoS predicts divergence of pressure at a glassy or jammed state—an important feature absent in the equations of state derived from existing PEL models [15,16].

# The PEL formalism

Within the PEL framework, the partition function of the system can be written as [3,6,7]:

$$Q(N,T,V) = \int exp\{-\beta[\epsilon_{is} - TS_c(N,V,\epsilon_{is}) + F_{vib}(N,T,V,\epsilon_{is})]\}d\epsilon_{is} \quad (1)$$

where $N$ is the total number of particles, $T$, temperature, $V$, total volume, $\beta = 1/k_B T$, $k_B$, the Boltzmann constant, $S_c$, the configurational entropy, $F_{vib}$, vibrational free energy, and finally, $\epsilon_{is}$ is the potential energy (a local property) for a given inherent structure, or basin. An apparent advantage of the PEL theory is that it simplifies the complex task of evaluating the partition function in a multidimensional phase space into the computation of a single integral in one-dimensional energy space [3,5]. The key component of the theory, Eq.(1), is the configurational entropy.

For calculating the configuration entropy, $S_c$, one starts with the local potential energy distribution of the inherent structures ($\epsilon_{IS}$) at a given temperature and volume. Currently, the Gaussian distribution is mostly adopted [6-13]:

$$\Omega(\epsilon_{is}; \sigma, E_0, c_0) = \frac{e^{c_0 N}}{\sigma\sqrt{2\pi}} exp\left[-\frac{1}{2}\left(\frac{\epsilon_{is} - \epsilon_0}{\sigma}\right)^2\right] \quad (2)$$

where $\epsilon_0$ is the mean (expected) value of the energy, $\sigma$, the variance, or the scaling parameter, the quantity, $e^{c_0 N}$, accounts for the total number of the inherent structures, or basins. Then the configurational entropy is given by:

$$S_c(N,V,\epsilon_{is})/k_B = ln\frac{e^{c_0 N}}{\sigma\sqrt{2\pi}} - \frac{1}{2}\left(\frac{\epsilon_{is} - \epsilon_0}{\sigma}\right)^2 \quad (3)$$

The local vibrational free energy $F_{vib}(N,T,V,\epsilon_{is})$ is usually assumed to be a linear function of the potential energy, $\epsilon_{is}$ [6,7]:

$$F_{vib}(N,T,V,\epsilon_{is}) = a_v + b_v \epsilon_{is} \quad (4)$$

where the parameters $a_v$ and $b_v$ are volume (density)-dependent. In the PEL formalism, the system in equilibrium samples a narrow range of $\epsilon_{is}$, and one adopts a saddle point approximation at $\epsilon_{is} = E_{IS}$, which is determined from the lowest free energy constraint, $\partial F(N.V,T;\epsilon_{is})/\partial \epsilon_{is} = 0$, with $\beta F(N.V,T;\epsilon_{is}) = -lnQ(N,T,V)$. Or equivalently [11,12],

$$1 - T\left[\frac{\partial S_c(N,V,\epsilon_{is})}{\partial \epsilon_{is}}\right]_{N,V} + \left[\frac{\partial F_{vib}(N,V,\epsilon_{is})}{\partial \epsilon_{is}}\right]_{N,V,T} = 0 \quad (5)$$

Eq.(3), Eq.(4) and Eq.(5) lead to the temperature dependence from the Gaussian distribution:

$$E_{IS} = E_0 - \frac{\sigma^2(1+b_v)}{2T} \quad (6)$$

Meanwhile the configurational entropy can be written as:

$$S_c = S_0 - \frac{1}{2}\left(\frac{E_{IS} - E_0}{\sigma}\right)^2 \quad (7)$$

where $S_0 = ln\frac{e^{c_0 N}}{\sigma\sqrt{2\pi}}$. Finally, with the saddle point approximation, Eq.(1) can be simplified as:

$$Q(N,T,V) \approx exp\{-\beta[E_{IS} - TS_c(N,V,E_{IS}) + F_{vib}(N,V,T,E_{IS})]\} \quad (8)$$

Hence the free energy can be written as:

$$F(N,V,T,E_{IS}) = (1+b_v)E_{is} - TS_c(N,V,E_{IS}) + a_v \quad (9)$$

Eq.(6), (7) and (9) comprise the PEL formalism with the Gaussian distribution. The parameters of the PEL theory are: $E_0$, $S_0$, $\sigma$, $b_v$, with $a_v$ being absorbed in $S_0$, and all the 4 parameters are volume (or density) -dependent.

The most appealing feature of this theory is its simplicity, and it does provide acceptable results for some supercooled fluids and glasses that exhibit strong or nearly strong characteristics. However, the Gaussian-distribution-based PEL theory has some notable shortcomings. As demonstrated below, this distribution can describe the low-temperature region for strong liquids and an intermediate-temperature region for fragile fluids. But it fails to capture the $E_{IS} \sim T$ relation over the wide range and is unable to describe the fragile-to-strong transition observed in various fluids [6-9].

The Gaussian distribution, as expressed in Eq. (2), assumes that the potential energy distribution of inherent structures is symmetric across the entire configurational space. Nevertheless, such a symmetric distribution holds for limited homogenous conditions and the deviation from a normal distribution becomes pronounced as the configurational space expands and the energy distribution becomes increasingly heterogeneous. Therefore, to accurately represent the reality, the potential energy distribution should be asymmetric, hence non-Gaussian. The Gamma distribution provides a suitable alternative for this purpose. With the PEL notations, Gamma distribution can be written as:

$$\Omega(\epsilon_{is}) = \frac{e^{cN}\lambda^\alpha}{\Gamma(\alpha)}\left(\frac{\epsilon_{is} - \epsilon_0}{\sigma_\epsilon}\right)^{\alpha-1} exp\left(-\lambda\frac{\epsilon_{is} - \epsilon_0}{\sigma_\epsilon}\right) \quad (10)$$

where $\alpha$ is the shape parameter; $1/\lambda$, the scalar parameter; the parameter $\sigma_\epsilon$ acts as the variance and the quantity $\frac{\epsilon_{is}-\epsilon_0}{\sigma_\epsilon}$ is dimensionless; $\Gamma(\alpha)$ is the Gamma function; $e^{cN}$, the total number of inherent structures. If we define $\sigma_+ = \sigma_\epsilon/\lambda$, Eq.(10) becomes:

$$\Omega(\epsilon_{is}) = \frac{e^{cN}\lambda}{\Gamma(\alpha)}\left(\frac{\epsilon_{is} - \epsilon_0}{\sigma_+}\right)^{\alpha-1} exp\left(-\frac{\epsilon_{is} - \epsilon_0}{\sigma_+}\right) \quad (11)$$

By using Eq.(4), Eq.(5) and Eq.(11), or using the saddle point approximation, we have:

$$\frac{S_c}{k_B} = S_0 + (\alpha - 1)ln\left(\frac{E_{IS} - E_0}{\sigma_+}\right) - \frac{E_{IS} - E_0}{\sigma_+} \quad (12)$$

where $S_0 = ln\frac{e^{cN}\lambda}{\Gamma(\alpha)}$, and the temperature dependence of $E_{IS}$ is given by:

$$E_{IS} = E_0 + \frac{aT}{b+T} \quad (13)$$

where $a = (\alpha - 1)\sigma_1$, $b = \sigma_1(1 + b_v)$. The free energy is given by Eq.(9). The parameters of the Gamma-distribution based PEL formalism are $S_0$, $\alpha$, $E_0$, $\sigma_+$, $b_v$. One additional parameter, $\alpha$, is introduced with the Gamma distribution, compared with the Gaussian distribution. As $\alpha \to$ 1, the distribution becomes exponential, and as $\alpha \to \infty$, the distribution becomes Gaussian-like (See Figure 1).

Eq.(13) warrants further discussions. If we expand the equation, $E_{IS} = E_0 + a\left[1 - \frac{b}{T} + \left(\frac{b}{T}\right)^2 - \left(\frac{b}{T}\right)^3 + \cdots\right]$ and omit higher order (>=2) terms, then Gaussian relation, Eq.(6), is recovered. As $T \to 0$, $E_{IS} \to E_0$, therefore for crystalline solid, $E_0$ is the lowest energy the system can possess. For a glassy system, we can adopt the concept of effective temperature [6], then rewrite Eq.(13) as $E_{IS} = E_0 + aT_{eff}/(b + T_{eff})$, hence the state at $E_0$ can be seen as a glassy or jammed state with $T_{eff} \to 0$. But in the following calculations, we will only use $T$, not $T_{eff}$.

In discussing glass transition, an important concept, the Kauzmann paradox, needs to be addressed. The so-called Kauzmann point [13] is determined by the zero-configuration entropy, $S_c = 0$. For the Gamma distribution, this is readily achievable. From Eq.(12), after some algebra, we obtain

$$E_K = E_0 - \sigma_+(\alpha - 1)W\left(-\frac{1}{\alpha-1}e^{-\frac{S_0}{\alpha-1}}\right) \quad (14)$$

where $W$ is the Lambert W(x) function. Keep in mind that in the domain of $\{-\frac{1}{e} \leq x < 0\}$, there exist two branches for the W function: $W_0$, $W_{-1}$. As $a > 0$ (fragile liquid) $W_0$ should be used, while as $a < 0$ (strong liquid), $W_{-1}$ should be adopted. If Eq.(14) is inserted into Eq.(13), one gets the Kauzmann temperature:

$$T_K = \frac{b(E_K - E_0)}{a - (E_K - E_0)} \quad (15)$$

Finally, from Eq.(4), Eq.(9) and Eq.(12), we can rewrite the free energy as:

$$F(N,V,T;E_{IS}) = F_0 - k_BT(\alpha - 1)ln\left(\frac{E_{IS} - E_0}{\sigma_+}\right) + g\frac{(E_{IS} - E_0)}{\sigma_+} \quad (16)$$

where $F_0 \equiv -k_B T S_0 + (1+b_v)E_0$, $g \equiv [k_B T + a_v + (1+b_v)\sigma_+]$, and the equation of state (EoS) can be derived:

$$Z = \frac{P}{\rho k_B T} = \beta \rho \frac{\partial F_0}{\partial \rho}$$
$$-\rho ln\left(\frac{E_{IS}-E_0}{\sigma_+}\right)\frac{\partial \alpha}{\partial \rho} +$$
$$\frac{\rho}{k_B(b+T)}\left[\frac{\partial g\alpha}{\partial \rho}+a(\alpha-1)\frac{\partial b}{\partial \rho}\right]$$
$$-\frac{g a \rho}{k_B(b+T)^2}\frac{\partial b}{\partial \rho} - k_B^{-1} a(\alpha-1)\rho\frac{\partial a}{\partial \rho} \quad (17)$$

where $Z$ is the compressibility. In Eq.17) we assume that $V$ is the molar volume and $\rho = 1/V$, the density. In high density region, $\frac{\partial \alpha}{\partial \rho} > 0$, by mathematical relation, $\lim_{x\to 0^+} ln(x) \to -\infty$, therefore, a singularity is included in the EoS:

$$-\lim_{E_{IS}\to E_0} ln\left(\frac{E_{IS}-E_0}{\sigma_+}\right) \to \infty \quad (18)$$

Namely, as $E_{IS} \to E_0$, a divergent term is resulted when the system is "trapped" in a glassy or jammed at state $E_0$. For a crystalline solid, Eq.(18) is equivalent to $T \to 0$. For supercooled liquid or glasses, it is equivalent to $T_{eff} \to 0$, as mentioned. As discussed below, $E_0$ should be treated differently for fragile and strong liquids, but for convenience, we use one notation, $Z_J$, for both. With the similar arguments discussed in refs [15,16], we have a new type of EOS for supercooled liquid and glasses:

$$Z = Z_{IS} + Z_{vib} + Z_J \quad (19)$$

where $Z_{IS}$ is the from the IS contribution ($\alpha$, $\sigma_+$, $S_0$, $E_0$), $Z_{vib}$ from vibration contribution ($a_v$, $b_v$) and the component for the "jammed" state is defined as

$$Z_J = -\rho ln\left(\frac{E_{IS}-E_0}{\sigma_+}\right)\frac{\partial \alpha}{\partial \rho} \quad (20)$$

This term is required for an EoS to be applicable to the entire density range since $Z_J \to \infty$ as $\rho \to \rho_J$, where $\rho_J$ is the density at the jammed state. For example, a widely used EoS for supercooled liquids and glasses is $Z_J \propto (\rho_J - \rho)^{-1}$ [17,18].

In addition, Eq.(17) contains terms with $(b+T)^{-1}$ and $(b+T)^{-2}$. As shown below, for strong fluids, $b < 0$, therefore at $T = -b$, the pressure diverges, $Z \to \infty$. E.(13) indicates that this corresponds to $E_{IS} \to -\infty$ (since $a < 0$). Consequently, the system becomes trapped at a glassy state, $T_g = -b$.

Before applying the revised PEL theory to real fluids, we first compare the Gamma distribution, Eq.(10), with the Gaussian distribution, Eq.(2), using some numerical results, as illustrated in Figure 1. For supercooled liquids, the Gamma distribution is skewed towards the lower energy side. As the shape factor increases, $\alpha \to \infty$, the distribution gradually becomes Gaussian-like. In the low $E_{IS}$ region, the Kauzmann point ($S_c = 0$) is generated, similar to the Gaussian model (dotted lines). However, in the high $E_{IS}$ region, the unphysical point (the second $S_c = 0$ point generated by the Gaussian distribution) does not emerge within the range of interest.

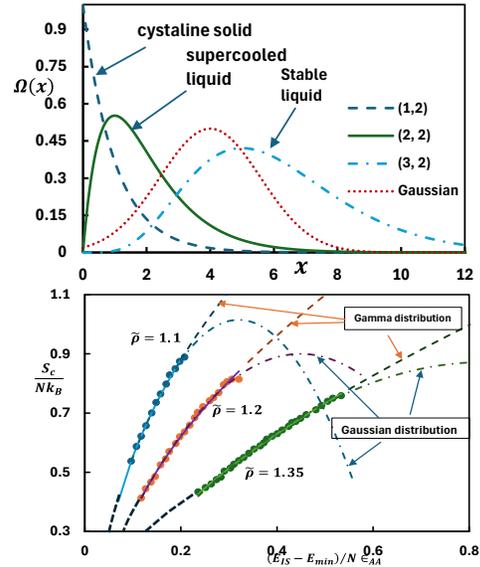

**Figure 1**. Gamma distribution vs Gaussian distribution. Above: probability density function of Gamma distribution vs Gaussian distribution, the number pairs in parenthesis are $(\alpha, \theta = 1/\lambda)$. Below:

configurational entropy for BMLJ [13] at several densities. The calculation details for Gaussian distribution (dashed lines) can be found from ref. [13]. For the Gamma distribution, Eq.(12) is used with $(E_{IS} - E_0)/\sigma_+ = c_1 (E_{IS} - E_{min})/N\epsilon_{AA}$, and the fitting parameters are $S_0, (\alpha - 1)$ and $c_1$: at $\tilde{\rho} = 1.1$, $S_0 = 1.2427$, $\alpha - 1 = 0.3402$, $c_1 = -0.9256$; at $\tilde{\rho} = 1.2$, $S_0 = 0.91154$, $\alpha - 1 = 0.2740$, $c_1 = -0.75281$; at $\tilde{\rho} = 1.35$, $S_0 = 0.61654$, $\alpha - 1 = 0.21955$, $c_1 = -0.53694$.

## Applications

The Gamma-distribution based PEL theory has been applied to three systems for which simulation data for configurational entropy, $S_c$, and/or potential energy, $E_{IS}$, are available. The systems include (a) a binary mixture of the Lennard-Jones fluid (BMLJ) [6,13,14] with van-der-Waals type interactions; (b) a network-forming system, SiO2 [8,9,10]; (c) q-TIP4P/F water [11,12] with hydrogen-bonds. For each system, 5 parameters are determined by fitting the data for $S_c$ and $E_{IS}$ simultaneously if available.

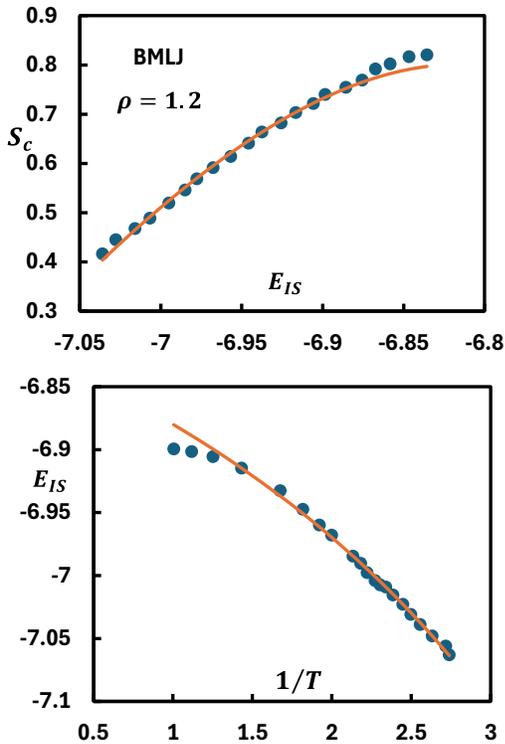

**Figure 2. BMLJ fluid**. Above: $S_c \sim E_{IS}$; below: $E_{IS} \sim 1/T$. Solid lines are calculated from Eq.(12) and Eq.(13), respectively. Data points are from ref. [14]. The fitted values of parameters: $\alpha - 1 = 3.2636$, $S_0 = 0.2049$, $E_0 = -6.4347$, $a = -0.38109$, $b = -0.14383$, $\sigma_1 = a/(\alpha - 1)$. The Kauzmann constants calculated by Eq.(14) and Eq.(15), respectively: $E_K = -7.138$, $T_K = 0.3139$.

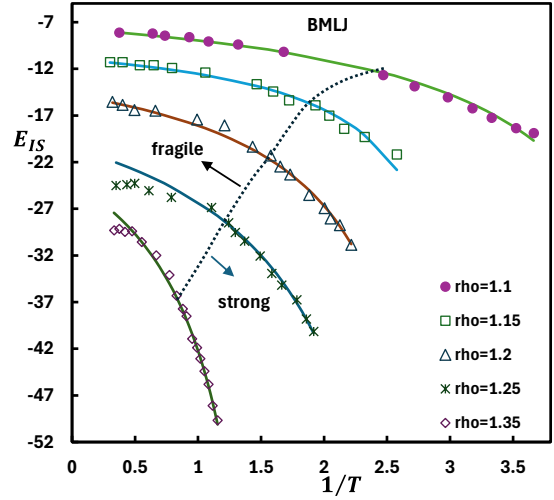

**Figure 3**. $E_{IS} \sim 1/T$ relation for the MBLJ fluid at various densities. The data points are reproduced from ref.[13] (Figure 3C of the ref. with the parameters provided in the caption). The solid lines are calculated by Eq.(13).

Figure 2 presents the calculation results for the BMLJ fluid, illustrating the $S_c \sim E_{IS}$ relation and $E_{IS} \sim 1/T$ relation, respectively. Figure 3 depicts $E_{IS} \sim 1/T$ relation for the BMLJ fluids at several densities. A dashed line is used to divide the entire liquid region into regimes for fragile and strong liquid. The Gamma-distribution based PEL theory can describe both $S_c$ and $E_{IS}$ over the entire temperature range, except few points at high temperature end. In contrast, the Gaussian-distribution based PEL theory for the $E_{IS} \sim 1/T$ relation is only applicable in the low temperature region. The values of the parameters are provided in Table S1 (the Supplementary Information, SI).

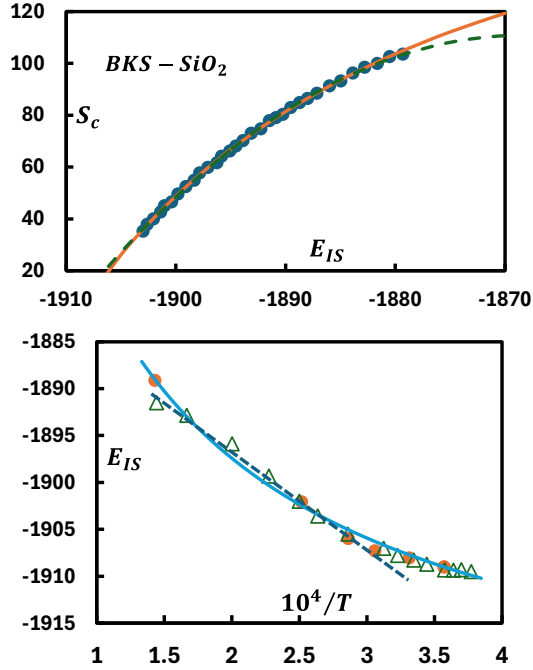

**Figure 4.** Configurational entropy and inherent structure energy for BKS $SiO_2$ at $\rho = 2.36\ g/cm^3$. Points are simulation data from refs.[8,9,10]. Solid lines are from Eq.(12) and Eq.(13), respectively, $\alpha-1 = 153.29$, $S_0 = -467.56$, $E_0 = -1919.01$, $a = 120.46$, $b = 1.4069 \times 10^4$. Dashed lines are from the Gaussian distribution.

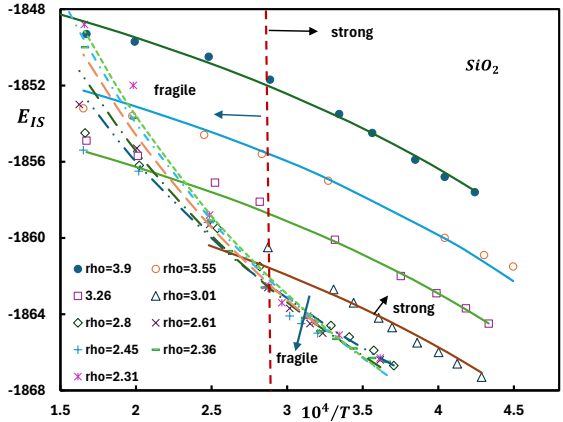

**Figure 5.** $E_{IS} \sim 1/T$ relation for BKS $SiO_2$ at different densities. Simulation data points are from Refs.[8,9, 10]. Lines are calculated by Eq.(13). Solid lines are for strong or partially strong fluids; dot-dashed lines are for fragile fluids.

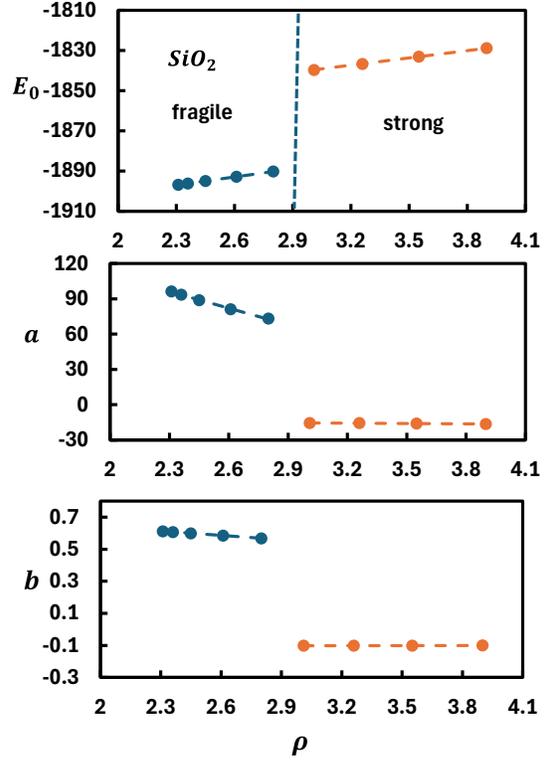

**Figure 6.** Density dependence of the parameters of Eq.(13) for BKS silica. The numerical values are listed in Table S2 (SI).

Figure 4 to Figure 6 present the calculation results for BKS Silica. Silica melts are well known network-forming liquids, and usually characterized as strong fluids. However, over a specific density region they exhibit a clearly fragile-to-strong crossover [13]. At density $\rho < 3.0$ the fluids are fragile, and the Gaussian distribution fails to describe the behavior in the low temperature region. At high densities, the fluids show a Gaussian or partial-Gaussian distribution feature. Figure 5 illustrates the division between the fragile and strong fluids. In the entire area to left of the dotted line and below the solid line for $\rho = 3.0$, the system exhibits fragile features, while in the remaining area (the upper-right portion) it displays strong features. Figure 6 depicts the fitting parameters with Eq.(13). A notable observation is the discontinuous in parameter values. For parameters $a$ and $b$, the signs change. For fragile regime, $a > 0$, $b > 0$, $\partial^2 E_{IS}/\partial T^2 > 0$, indicating the $E_{IS} \sim 1/T$ curve is concave up (convex). In contrast, for strong fluids, $a < 0$, $b < 0$, , $\partial^2 E_{IS}/\partial T^2 < 0$, $E_{IS} \sim 1/T$ indicating a concave down (concave).

Next, the Gamma-distribution-based PEL theory is applied to a unique liquid characterized by hydrogen bonding: q-TIP4P/F water. Simulation data on configurational entropy and inherent structure energy have been reported over broad ranges of density and temperature [11, 12].

Figure 7a and Figure 7b present the results for $S_c \sim 1/T$ and $E_{IS} \sim 1/T$ relations, respectively. The predicted relations for $S_c \sim E_{IS}$ at different densities are shown by Figure 8. The accuracies of the predictions are guaranteed by the fitting accuracies shown in Figure 7. As expected, at low densities the values of the configurational entropy are higher than those at high densities. The values of the fitting parameters are listed in Table S3 (SI) and their plots are also shown in the SI.

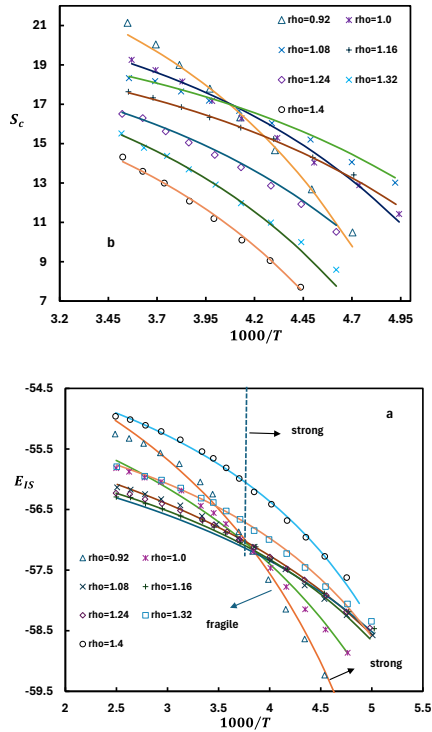

**Figure 7**. Fitting results for $S_c \sim 1/T$ (a) and $E_{IS} \sim 1/T$ (b). Data points are from Ref.[11].

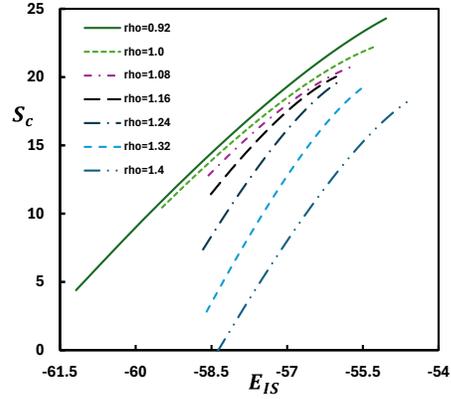

**Figure 8**. Configurational entropy vs inherent structure energy for q-TIP4P/F water system predicted with Gamma-distribution based PEL theory.

In the density range considered, the q-TIP4P/F water system exhibits characteristics of a partially strong liquid: at low temperature region, the Gaussian distribution applies, while it begins to fail at higher temperatures (Figure 7b). This behavior is similar to that observed in the SiO2 system (Figure 5) and the BMLJ system (Figure 3). Consequently, we conjecture that the fragile-to-strong transition, or polyamorphism, is a universal phenomenon for all fluids. This suggests that distinguishing different molecular fluids based on fragility (strong or fragile) may not be necessary for their characterization.

As shown by Figure S1 (Supplementary Information, SI), the fitting parameters can be effectively correlated with density using simple polynomial functions. The values of the generalized parameters are listed in Table S4. This is particularly useful: utilizing these correlations, as expressed in Eq. (16) and Eq. (17), enables the derivation of a reliable equation of state (EoS), which facilitates calculations of various thermodynamic properties, as discussed in Ref. [11]. Further applications will not be addressed in this work.

## Discussions and conclusions

From a statistical perspective, with the Gaussian distribution a quantity $x$ randomly varies about a mean value $x_0$. Due to the symmetry of the squared term, $(x - x_0)^2$, the contribution to the probability density is the same regardless the approach is from the left ($x < x_0$) or from the right ($x > x_0$). This symmetric behavior makes the Gaussian distribution inadequate for describing the fragility change of a supercooled liquid. In 1980's, a skew Gaussian distribution [19], $f(x) = 2\phi(x)\Phi(\alpha_s, x)$, has been proposed for modifying the Gaussian distribution to account for the asymmetric shifts, where $\Phi(\alpha_s, x)$ is related to the error function and $\alpha_s$ is the extra parameter accounting for the skewness of the distribution. The detailed application of the skew Gaussian distribution to supercooled liquids and glasses is presented in the SI. Since the error function is involved, an explicit solution for $E_{IS}$ is infeasible. Here, we provide only a brief discussion to aid in understanding the nature of the fragility transition.

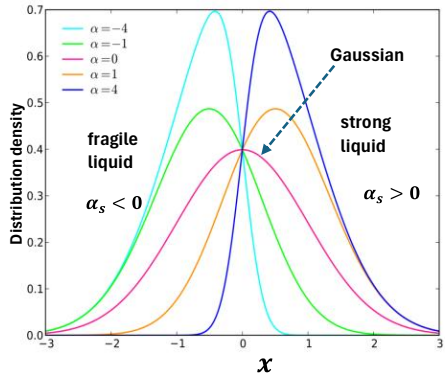

**Figure 9**. The skew Gaussian distribution compared to the Gaussian distribution (middle), partially adapted from ref.[19].

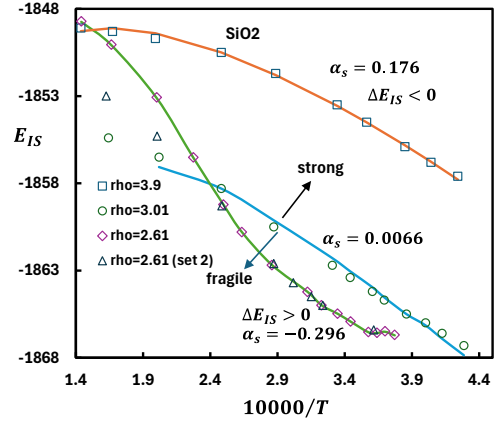

**Figure 10**. The $E_{IS} \sim 1/T$ relation form the skew Gaussian distribution applied to the SiO2 system.

Figure 9 compares the Skew Gaussian distribution with the standard Gaussian distribution. As $\alpha_s < 0$, the distribution is left-skewed, while for $\alpha_s > 0$, right-skewed. Figure 10 depicts some calculation results for SiO2 system discussed previously, where $\Delta E_{IS} = E_{IS} - E_0$. Some important conclusions can be drawn from these results. For the left-skewed case, $\alpha_s < 0$, $\Delta E_{IS} > 0$, indicating that the system at the state $E_{IS}$ is less "stable" compared to the state at $E_0$, making it fragile. For the right-skewed case, $\alpha_s > 0$, $\Delta E_{IS} < 0$, indicating that the system at the state $E_{IS}$ is more "stable", characterizing the system as strong.

Returning to the Gamma distribution case. Figure 11 illustrates how $\Delta E_{IS}$ changes with temperature for the SiO2 system for a low density, $\rho = 2.8$ and a high density, $\rho = 3.26$, respectively. For the case at $\rho = 2.8$, $\Delta E_{IS} > 0$, indicating the system is fragile, which corresponds to a left-skewed Gaussian distribution. In contrast, for the case at $\rho = 3.26$, $\Delta E_{IS} < 0$, signifying that the system is strong, corresponding to a right-skewed Gaussian distribution. Thus, the Gamma distribution effectively captures the fundamental characteristics of the skew Gaussian distribution, in consistent with the Skew Gaussian distribution.

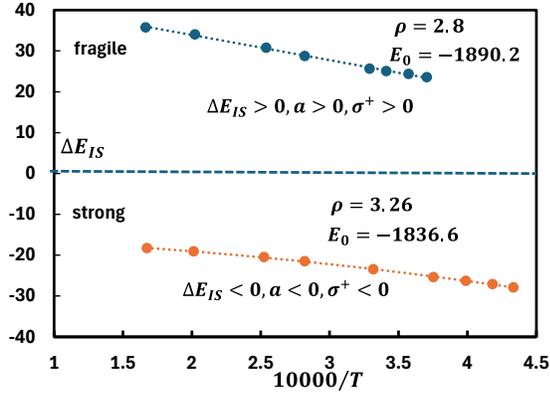

**Figure 11**. $\Delta E_{IS} \sim 1/T$ relation for SiO2 system at two densities.

In conclusion, in this work we developed a PEL theory by utilizing the Gamma distribution to represent the configurational entropy. The proposed model, which includes five parameters, effectively captures the configurational entropy and the potential energy of inherent structures over wide ranges of temperature and density. In comparison, the Gaussian distribution is limited to specific cases: (1) strong liquids in the low-temperature region and (2) fragile liquids in a limited intermediate-temperature region. The new Gamma-distribution-based PEL theory also provides analytical expressions for the Kauzmann constants, offering a more comprehensive framework for investigating the thermodynamic properties of supercooled liquids and glasses.

For the first time, it is demonstrated that the transition of from fragile to strong (or vice versa) is associated with a shift in the potential energy distribution and with the curvature changes of the $E_{IS} \sim 1/T$ relation, characterized by the discontinuities of the model parameters.

Another interesting observation (Figure 6) is from values of $E_0$: $E_{0,strong} > E_{0,fragile}$. For a fragile supercooled liquid to transition to a strong one, an energy input (about 2%) is required. Or as a strong liquid transition to fragile one some energy will be released. Given the small magnitude of this energy change, the fragile-to-strong transition can be viewed as analogous to a (weak) liquid-liquid transition.

Within the potential energy ($E_{IS}$) ranges under discussion, the state at $E_{0,fragile}$ corresponds to the lowest energy for fragile liquid, $\Delta E_{IS} > 0$, while for strong fluid $E_{0,strong}$ exhibits the highest value, $\Delta E_{IS} < 0$. According to the EoS, Eq.(17), for a fragile fluid the system at $E_0$ can be seen as "glassy" and the pressure diverges. For a strong fluid, the system at $E_{IS} \to \infty$ (or $T = -b$), is glassy; while at $E_0$ the system can be seen as "jammed", and the pressure diverges. Such a classification is left open for further discussions.

### Supplementary information

In this Supplementary Information (SI), some calculation details and values of parameters are provided.

**Table S1.** The parameters for BMLJ fluid (Figure 3)

| $\rho$ | $E_0$ | $a$ | $b$ |
|---|---|---|---|
| 1.35 | -9.85988 | -9.54690 | -0.62727 |
| 1.25 | -8.0645 | -12.50999 | -0.32072 |
| 1.2 | -7.48412 | -6.81903 | -0.31928 |
| 1.15 | -7.11528 | -3.83307 | -0.29325 |
| 1.1 | -6.95798 | -5.53639 | -0.18676 |

**Table S2** SiO2 parameters (Figure 5)

| $\rho$ | $E_0$ | $a$ | $b/10000$ |
|---|---|---|---|
| 2.31 | -1896.88 | 96.2434 | 0.61226 |
| 2.36 | -1896.2 | 93.5136 | 0.60774 |
| 2.45 | -1894.98 | 88.8139 | 0.5996 |
| 2.61 | -1892.82 | 81.1379 | 0.58514 |
| 2.8 | -1890.25 | 73.1517 | 0.56796 |
| 3.01 | -1839.64 | -15.539 | -0.1011 |
| 3.26 | -1836.6 | -15.702 | -0.1008 |
| 3.55 | -1833.08 | -16.003 | -0.1004 |
| 3.9 | -1828.83 | -16.526 | -0.1 |

**Table S3**. The values of the IS parameters and Kautzmann constants for q-TIP4P/F water

| $\rho$ | $\alpha - 1$ | $S_0$ | $E_0$ | $a$ | $b/1000$ | $T_K$ | $E_K$ |
|---|---|---|---|---|---|---|---|
| 1.4 | 19.20269 | -18.2181 | -51.7005 | -2.11153 | -0.13681 | 200.0776 | -58.3781 |
| 1.32 | 19.10626 | -16.9972 | -52.9472 | -1.87302 | -0.1337 | 193.313 | -59.0212 |
| 1.24 | 18.73963 | -15.6746 | -53.3484 | -1.97735 | -0.12565 | 180.4595 | -59.859 |
| 1.16 | 18.35525 | -14.4011 | -53.0669 | -2.30534 | -0.11534 | 164.6717 | -60.7621 |
| 1.08 | 18.20558 | -13.3274 | -52.2655 | -2.73781 | -0.11301 | 160.0623 | -61.5798 |
| 1.0 | 18.54308 | -12.6042 | -51.1072 | -3.1556 | -0.1246 | 174.3812 | -62.1611 |
| 0.92 | 19.62019 | -12.3821 | -49.7546 | -3.43953 | -0.1398 | 192.5964 | -62.3017 |

For q-TIP4P/F water, the density dependence of IS parameters can be expressed with polynomial function:

$$Y = \sum_{k=0}^{n} c_k \rho^k \qquad (S1)$$

For example, as $Y = \sigma - 1, n = 3$:

$$\sigma - 1 = -82.1793\rho^3 + 304.3205\rho^2 - 370.419\rho + 166.8205 \qquad (S2)$$

The values of the coefficients for all the parameters are listed in Table S4. Figure S1 depicts the density-dependence curves for all IS parameters. Figure S2 shows density dependence for the Kauzmann constants, $E_K$ and $T_K$.

**Table S4**. Values of coefficients in generic correlation, Eq.(S1)

| $k/c_k$ | $\alpha - 1$ | $S_0$ | $E_0$ | $a$ | $b$ |
|---|---|---|---|---|---|
| 5 | 0 | 0 | 0 | 0 | -30.2537 |
| 4 | 0 | 0 | 0 | 0 | 179.8715 |
| 3 | -82.1793 | 49.05421 | 53.00463 | -38.7935 | -423.909 |
| 2 | 304.3205 | -186.316 | -143.842 | 126.8382 | 494.5708 |
| 1 | -370.419 | 219.2471 | 112.6378 | -132.662 | -285.418 |
| 0 | 166.8205 | -94.5896 | -72.9074 | 41.46173 | 65.0139 |

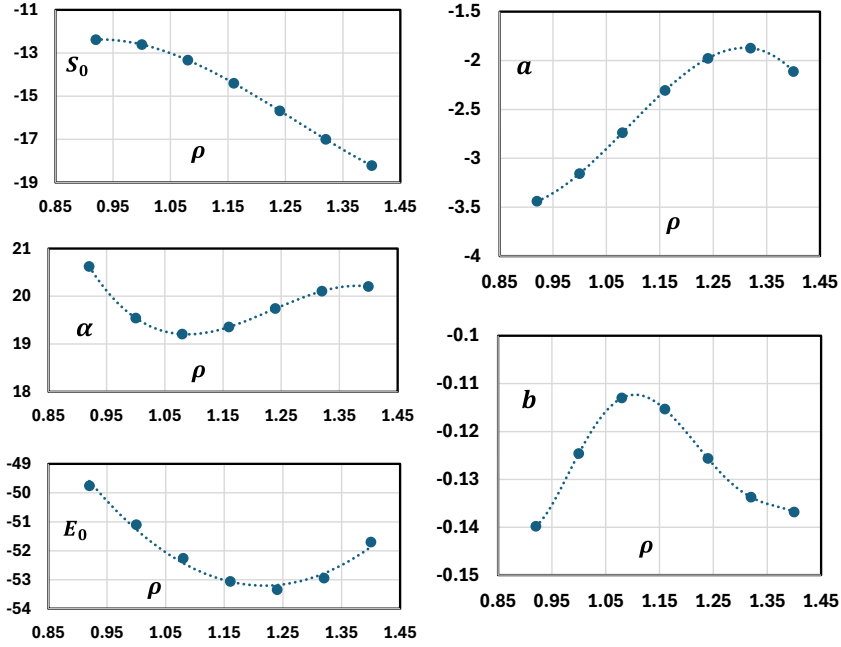

**Figure S1**. Plots of IS parameters vs density.

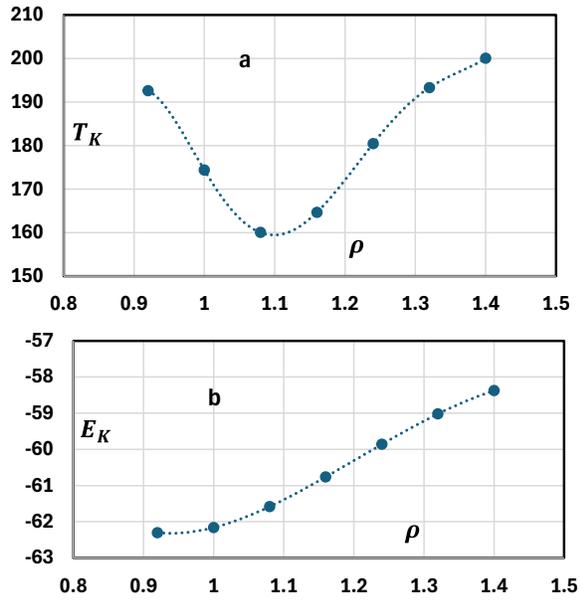

**Figure S2**. Plots of Kauzmann constants vs density.

**Skew Gaussian distribution**

For our current PEL application, the skew Gaussian distribution is written as:

$$\Omega = \frac{1}{\sigma\sqrt{2\pi}} exp\left[-\frac{1}{2}\left(\frac{\epsilon_{is}-\epsilon_0}{\sigma}\right)^2\right]\left\{1+\text{erf}\left[\frac{1}{\sqrt{2}}\alpha_0\left(\frac{\epsilon_{is}-\epsilon_0}{\sigma}\right)\right]\right\} \qquad (S3)$$

let $\alpha_s = \alpha_0/2$

$$\Omega = \frac{1}{\sigma\sqrt{2\pi}} exp\left[-\frac{1}{2}\left(\frac{\epsilon_{is}-\epsilon_0}{\sigma}\right)^2\right]\left\{1 + \text{erf}\left[\alpha_s\left(\frac{\epsilon_{is}-\epsilon_0}{\sigma}\right)\right]\right\} \quad (S4)$$

The configurational entropy is given by:

$$S_c(E_{is}) = ln\Omega = -ln(\sigma\sqrt{2\pi}) - \frac{1}{2}\left(\frac{\epsilon_{is}-\epsilon_0}{\sigma}\right)^2 + ln\left\{1 + \text{erf}\left[\alpha_s\left(\frac{\epsilon_{is}-\epsilon_0}{\sigma}\right)\right]\right\} \quad (S5)$$

The free energy reads:

$$F(T,V;\epsilon_{is}) = -\frac{1}{\beta}S_c(\epsilon_{is}) + (1+b_v)\epsilon_{is} + a_v \quad (S6)$$

Or

$$F(T,V;\epsilon_{is}) = \frac{1}{\beta}ln(\sigma\sqrt{2\pi}) + \frac{1}{\beta}\frac{1}{2}\left(\frac{\epsilon_{is}-\epsilon_0}{\sigma}\right)^2 - \frac{1}{\beta}ln\left[1 + \text{erf}\left(\alpha_s\frac{\epsilon_{is}-\epsilon_0}{\sigma}\right)\right] + (1+b_v)\epsilon_{is} + a_v \quad (S7)$$

For applying Eq.(5):

$$\frac{\partial F(T,V;\epsilon_{is})}{\partial \epsilon_{is}} = \frac{1}{\beta}\frac{1}{\sigma}\left(\frac{\epsilon_{is}-\epsilon_0}{\sigma}\right) - \frac{1}{\beta}\frac{\frac{2}{\sqrt{\pi}}\frac{\alpha_s}{\sigma}exp\left[-\left(\alpha_s\frac{\epsilon_{is}-\epsilon_0}{\sigma}\right)^2\right]}{1 + \text{erf}\left(\alpha_s\frac{\epsilon_{is}-\epsilon_0}{\sigma}\right)} + 1 + b_v = 0 \quad (S8)$$

Then at $\epsilon_{is} = E_{is}$, $\epsilon_0 = E_0$, we have:

$$\alpha\left(\frac{E_{is}-E_0}{\sigma}\right) - \frac{\frac{2}{\sqrt{\pi}}\sigma\alpha_s\frac{\alpha_s}{\sigma}exp\left[-\left(\alpha_s\frac{E_{is}-E_0}{\sigma}\right)^2\right]}{1 + \text{erf}\left(\alpha_s\frac{E_{is}-E_0}{\sigma}\right)} + \sigma\alpha_s(1+b_v)\beta = 0 \quad (S9)$$

Define:

$$x = \alpha_s\frac{E_{is}-E_0}{\sigma} = \Delta E_{is}\frac{\alpha_s}{\sigma}$$

The configurational entropy is given by:

$$S_c(E_{is}) = S_0 - \frac{1}{2}\left(\frac{x}{\alpha_s}\right)^2 + ln[1 + \text{erf}(x)] \quad (S10)$$

And finally, the $E_{is}\sim T$ relation is given by an implicit equation:

$$x - \alpha_s^2\frac{\frac{2}{\sqrt{\pi}}e^{-x^2}}{1 + \text{erf}(x)} + \sigma\alpha_s(1+b_v)\frac{1}{T} = 0 \quad (S11)$$

For SiO2, the parameters used for Figure 10 are listed in Table S5.

**Table S5**. Values of parameters of the skew Gaussian distribution for SiO2

| $\rho$ | $b_v$ | $\alpha_s$ | $E_0$ | $\sigma$ |
|---|---|---|---|---|
| 3.90 | -0.8086 | 0.1761 | -1681.1 | 7.8922 |
| 3.01 | -0.2997 | 0.00663 | -1641.0 | 0.27248 |
| 2.61 | -0.8839 | -0.2962 | -1929.9 | 3.0730 |

**The Lambert W function.**

For fragile fluid, the Lambert W function takes the $W_0$ value, and can be calculated with the Excel function, LAMBERTW(x). For strong fluids, the Lambert W(x) function takes the $W_{-1}$ value, and can be calculated with the following approximation equation (x values within the interval, $(-1/e, 0)$:

$$L_1 = ln(-x) \tag{S12}$$

$$L_2 = ln(-L_1) \tag{S13}$$

$$W_{-1} = L_1 - L_2 + \frac{L_2}{L_1} + \frac{L_2(-2 + L_2)}{2L_1^2} + \frac{L_2(6 - 9L_2 + 2L_2^2)}{6L_1^3} \tag{S14}$$

Usually, the first 4 terms are accurate enough for current purpose.